\documentclass{aa}  
\usepackage{graphicx}
\usepackage{amsmath}
\usepackage{enumitem}
\usepackage{booktabs}
\usepackage[normalem]{ulem}
\usepackage[table,xcdraw]{xcolor}
\usepackage{lineno}
\usepackage{subcaption}
\usepackage{lipsum}
\usepackage{acronym}
\acrodef{SDE}[SDE]{Stochastic Differential Equation}
\acrodef{GCR}[GCR]{Galactic Cosmic Ray}
\acrodef{CUDA}[CUDA]{Compute Unified Device Architecture}
\acrodef{CPI}[CPI]{Charged Particle Instrument}
\acrodef{CIR}[CIR]{Corotating Interaction Region}
\acrodef{HMF}[HMF]{Heliospheric Magnetic Field}
\acrodef{VLIS}[VLIS]{very local interstellar spectrum}
\acrodef{TPE}[TPE]{Transport Equation}
\acrodef{QTI}[QTI]{quiet-time increases}
\usepackage{txfonts}

\begin{document} 


\title{On the residence-time of Jovian electrons in the inner heliosphere}

   \author{A. Vogt\inst{1}
          \and
          N. E. Engelbrecht\inst{2}
          \and
          R. D. Strauss\inst{2}
          \and
          B. Heber\inst{1}
          \and
          A. Kopp\inst{2,3}
          \and
          K. Herbst\inst{1}
          }

   \institute{Institut f\"ur Experimentelle und Angewandte Physik, Christian-Albrechts Universit\"at zu Kiel,
Leibnizstra\ss e 11, D-24118 Kiel, Germany\\
              \email{vogt@physik.uni-kiel.de}
         \and
             Centre for Space Research, North-West University, 2520 Potchefstroom, South Africa
         \and
            Theoretische Physik IV, Ruhr-Universit\"at Bochum, Universit\"atsstr. 150, 44801 Bochum, Germany
             }

   \date{}

 
  \abstract
   {Jovian electrons serve as an important test-particle distribution in the inner heliosphere and have been used extensively in the past to study the (diffusive) transport of cosmic rays in the inner heliosphere. With new limits on the Jovian source function (i.e. the particle intensity just outside the Jovian magnetosphere), and a new set of in-situ observations at 1 AU for both cases of good and poor magnetic connection between the source and observer, we revisit some of these earlier simulations. }
   {We aim to find the optimal numerical set-up that can be used to simulate the propagation of 6 MeV Jovian electrons in the inner heliosphere. Using such a set-up, we further aim to study the residence (propagation) times of these particles for different levels of magnetic connection between Jupiter and an observer at Earth (1 AU). }
   {
   Using an advanced Jovian electron propagation model based on the stochastic differential equation (SDE) approach
   , we calculate the Jovian electron intensity for different model parameters. A comparison with observations leads to an optimal numerical set-up, which is then used to calculate the so-called residence (propagation) times of these particles.}
   {Comparing to in-situ observations, we are able to derive transport parameters that are appropriate to study the propagation of 6 MeV Jovian electrons in the inner heliosphere. Moreover, using these values, we show that the 
   method of calculating the residence time applied in former literature is not 
   suited to being interpreted as the propagation time of physical particles. This is due to an incorrect weighting of the probability distribution. We propose and apply a new method, where the results from each pseudo-particle are weighted by its resulting phase-space density (i.e. the number of physical particles that it represents). Thereby we obtain more reliable estimates for the propagation time.}
   {}


   \maketitle
%

\section{Introduction}

During the last decade, solving particle transport equations by means of \acp{SDE} became an increasingly popular tool due to increasing computational power. Whereas many of these studies are focused on \acp{GCR}, this work builds upon recent research regarding Jovian electrons \citep[see][and references therein]{vogt2018,Nndanganeni2018}. Since Jupiter is essentially a ``steady state point source'' of MeV electrons, Jovian electrons present an unique opportunity to study particle transport in the inner heliosphere. Detailed computational parameter studies have also become more feasible due to the enhanced computational capabilities acquired by utilizing Nvidia's \ac{CUDA} as described by \citet{Dunzlaff2015}. Here we study the residence time of energetic Jovians in the heliosphere \citep[see e.g.][and references therein]{McKibben-etal-2005}. Since Jupiter is assumed to release electrons continuously, these times are not directly measurable, and therefore have to be derived from theory/modelling. In order to develop a reliable estimation of the residence or propagation time, it is therefore necessary to 
\begin{enumerate}
    \item determine if the computational setup is realistic and to
    \item validate the transport parameters against spacecraft data.
\end{enumerate} 

This work will address both topics.

\begin{figure}
    \centering
    \includegraphics[width=0.9\columnwidth]{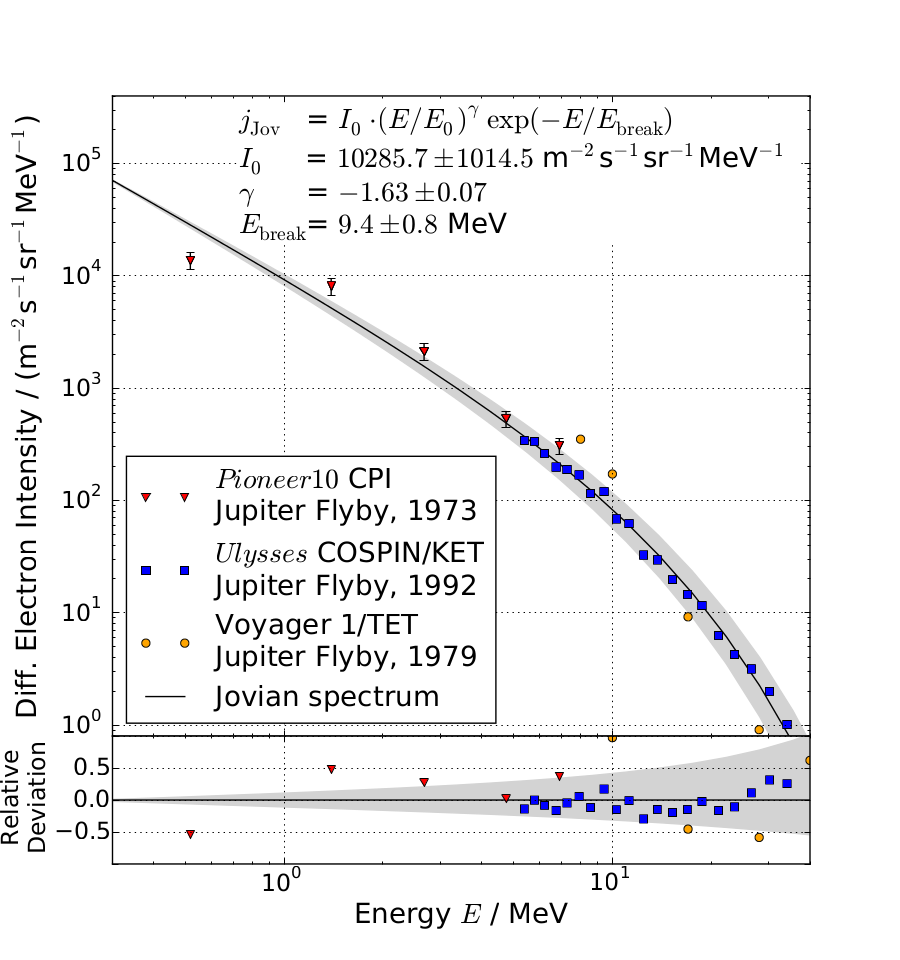}
    \caption{The Jovian source spectrum according to \cite{vogt2018}.The upper panel shows 
    the
    source spectrum as fitted to the \textit{Pioneer 10} CPI and \textit{Ulysses} COSPIN/KET data. The \textit{Voyager 1} TET data added in this plot seems to be in agreement as well. The lower panel complementary shows the relative deviation of the spacecraft data from the fit. The shaded area covers the $\pm\sigma$ uncertainty. }
    \label{fig:jovian_source_spectrum}
\end{figure}

As for modulation studies of \acp{GCR}, an essential constraint for such an investigation is the knowledge of the source spectrum. The Jovian electron source spectrum has been recently determined by \citet[][]{vogt2018} and is shown in Fig.~\ref{fig:jovian_source_spectrum}. Knowledge of this spectrum allows to estimate the effects of various physical parameters on computed Jovian intensities, which can also be compared to spacecraft observations of the same at Earth taken during periods of good and bad magnetic connection with the Jovian source. These comparisons can then provide insight as to the behaviour of quantities such as the low-energy electron diffusion coefficients parallel and perpendicular to the heliospheric magnetic field, as well as an optimised, realistic parameter set for further model computations 
as proposed in Tab.~\ref{tab:simulation_parameters}
, which is used to study the residence times of Jovian electrons. We focused our simulations on 6 MeV electrons during quiet-time conditions. This choice is in agreement with most prior investigations on Jovian electrons \citep[see e.g.][and references therein]{Kissmann2004, Nndanganeni2018} as it covers the detection range of several particle detection instruments such as Ulysses/KET, Voyager 1/TET, ISEE 3/ICE, IMP-8/CRNC and SOHO/EPHIN.

In order to estimate the residence times we propose a similar formalism as used to transform the probability densities resulting from the \ac{TPE} into differential intensities. As this is done by a convolution with the source spectrum \citep[see e .g.][and references therein]{Strauss2017}, an equivalent convolution is applied to the simulation times provided by the \acp{SDE} code. It is demonstrated that the method of calculating residence times employed by, e.g., \citet{Florinski2009} and \citet{Strauss2013}, cannot be interpreted as the propagation time of physical particles. A novel approach is proposed, where the results from each pseudo-particle are weighted by its resulting phase-space density (i.e. the number of physical particles that it represents), to obtain 
estimates for the propagation time 
that are more consistent with the limited observational constraints, but also more representative of the propagation time of the physical particles themselves.

\section{Scientific Background}

\subsection{Jovian Electrons as Test Particles}
\label{ssec:test_particles}

The scientific discussion of Jovian electrons dates back to the early 1970s when \cite{McDonald72} proposed the existence of a dominant Jovian source by addressing the correlation of the $\sim 13$ month periodicity in low-MeV electron counting rate measurements at Earth orbit with Jupiter's synodic period. After the Jupiter flyby of {\it Pioneer~10}, \citet{Teegarden1974} were able to confirm this hypothesis based on data obtained by the \ac{CPI}. \citet{Pyle1977} showed, by analysing the electron fluxes as function of distance and their dependence on the occurrence of \acp{CIR}, that the Jovian source is not only quasi-continuous but also point-like. 
This refers to the observation that no electron emission could be detected from Jupiter's magneto-tail which extends up to over 1 AU into the heliosphere
.   

As the dominant particle population from a few to several tens of MeV in the inner heliosphere, Jovian electrons soon became the subject of charged particle transport modelling \citep{Conlon1978,Zhang2017}. Due to Jupiter's decentral position, the magnetic connection by means of the Parker spirals determines whether the electrons reach the observer primarily via motion along the field or by diffusion perpendicular to it. Thus, Jovian electrons are ideal candidates to probe the electron diffusion coefficients in the inner heliosphere. 

Jovian electrons were used as test particles to model the charged-particle transport computationally \citep[see e. g.][and references therein]{Chenette77,Conlon1978,Fichtner2000,Ming07} to ascertain the  diffusion coefficients parallel and perpendicular to the \ac{HMF}. This was usually done by comparing computed with measured electron intensities at Earth during periods of good/bad magnetic connection. Furthermore, given the demonstrated sensitivity of computed low-energy galactic electron intensities to various turbulence quantities \cite[see][]{EB2010,EB2013,Engelbrecht2019}, it may be possible to draw conclusions from Jovian electrons as to the behaviour of those quantities in regions of the heliosphere where spacecraft observations of the 
same
do not exist \cite[see, e.g.,][]{Engelbrecht2017}. Since these transport parameters and 
the diffusion coefficients
depend on are highly spatially dependent, the time that particles reside in a certain part of the heliosphere may yield significant insights to the modulation of \acp{GCR} as well. \citet{Florinski2009} showed this dependency for \ac{GCR} protons, investigating the time they spend in the heliotail, the heliosheath and in the solar wind within the termination shock, respectively. Utilizing both galactic electrons and protons, 
\cite{strauss2011b}
focused on the connection between the total propagation time and energy losses. They find a significant non-linear dependency on the total propagation times, which is strong enough to influence also the observations of Jovian electrons.
These energy losses are entirely caused by adiabatic effects as other possible influences such as particle-particle interactions are 
negligible in
the \ac{TPE} due to a lack of significance in the interplanetary medium. As the adiabatic energy changes $dE/dt \sim -2/3\cdot Eu_{SW}/r$ 
are connected to the radial position, the corresponding energy loss rate per step only depend on the temporal step size $\Delta s$ and the radial position after the step. 
The radial direction of the step thereby is irrelevant. This leads to particles spending more simulation time at small radii losing more energy and implicitly to a statistical connection between the average energy losses and the particle's mean free paths. 

\subsection{The Jovian Electron Spectrum}
\label{ssec:jovian:source}

Parallel to the efforts of studying Jovian electron transport, the Jovian source itself has been investigated intensively. The main unknowns are the energy spectrum of the source and how the particles are accelerated. 

\begin{description}[wide=0\parindent, itemsep=1em]
\item[\bf Source energy spectrum:] Although widely considered to be dominant in its energy range, the shape and exact strength of the Jovian source remains a topic of debate. A first suggestion was published as soon as {\it Pioneer 10} confirmed the existence of the source by \cite{Teegarden1974}, but the limited amount of flyby data and the general difficulties inherent to measuring electrons, especially electron spectra, made it difficult to further constrain both the magnitude and the shape. Suggestions were published by \cite{Baker1976}, \cite{Eraker1982}, \cite{Haasbroek1997} and \cite{Ferreira2001} based on both {\it Pioneer 10}~CPI flyby and Earth orbit data. The two Voyagers and Ulysses are the only spacecraft equipped with particle instruments that could resolve the electron spectra above a few MeV \citep[see][and references therein]{Heber2005,vogt2018,Nndanganeni2018}. The latter two published source energy spectra $j_{jov}(E)$ shown in Fig.\ref{fig:jovian_source_spectrum} on the base of these flyby data. \citet{vogt2018} proposed 
\begin{equation}
j_{jov}(E)= \frac{1.029\cdot 10^{4}}{\mbox{m}^2\mbox{s sr MeV}} \left(\frac{E}{E_0}\right)^{-1.63} e^{-E/E_{b}}
\label{eqn:jovian_spectrum}
\end{equation}

with $E$, and $E_0$ the kinetic and rest energy of the electron, and $E_{b}=9.4$~MeV the spectral break energy. Whereas the exponent of $\alpha=-1.63$ is very much in agreement with the findings of \cite{Ferreira2001} and \cite{Baker1976}, the shape proposed in Eqn.~(\ref{eqn:jovian_spectrum}) is more similar to the suggestions by \cite{Teegarden1974} and \cite{Eraker1982} and acceleration theory in itself 
as \citet{Ferreira2001} proposed a combination of two spectra to fit the spectral break.
The {\it Voyager 1} flyby spectrum as published by \cite{Nndanganeni2018} is included in Fig.~\ref{fig:jovian_source_spectrum} and supports these results. An overview of the electron spectral data utilized for this study, obtained both during flybys and at Earth orbit, is listed in Tab.~\ref{tab:spectral_data}. Previous studies derived the spectra by solving the particle transport equation and fitting the results to measurements close to Earth \citep[see for example][and references therein]{Moses1987}.

\item[\bf Acceleration processes:] Since the findings of \cite{Bolton1989} it is widely accepted that the Jovian electrons originate in the solar wind, are picked up by the Jovian magnetosphere and diffuse inwards. {In situ} data has been obtained by spacecraft such as {\it Pioneer 10}, {\it Galileo} and {\it Cassini} and found to suggest acceleration via wave-particle interaction within the radiation belts, as discussed by \cite{Horne2008}, alongside adiabatic processes suggested by \cite{dePater1990}. Because the existence of the Jovian source seems to be linked to the planet's extended and strong magnetic field, the question as to whether these processes also apply close to Saturn is also relevant \cite[see e.g][]{Lange2008}. Recently, {\it Cassini} measurements \citep[as reported by][]{ Palmaerts2016,Roussos2016} revived this discussion and support, together with {\it Galileo} data, a dominant role of adiabatic processes within both planets' magnetospheres in order to accelerate MeV electrons \citep[see][and references therein]{Kohlmann2018}.
\end{description}

Due to the recent measurements of electrons outside the heliosphere \citep{Stone-etal-2013}, the question of how to distinguish the Jovian population from the Galactic background, based on suggestions for the \ac{VLIS} is discussed by \citet[][and references therein]{Bishoff-etal-2019}. Apart from its implications regarding the modulation of \acp{GCR}, \citet{Nndanganeni2018} find the Jovian population dominating the spectrum up to energies of about $E \sim 25~$MeV, followed by an energy range of $25~$MeV $\leq E \leq 40~$~MeV where the spectrum consists of a mixture of Jovian and Galactic electrons varied by the co-longitudinal dependency of the Jovian intensities. Regarding the radial dependency, Jovian electrons were found to be the dominant populations up to radial distances of $R \sim 15~$AU, based on simulations of the $6~$MeV Jovian and Galactic electron intensities. 

\begin{table*}
\centering
\caption{ Overview of the electron spectra used in this study.} 
\begin{tabular}{l l l}
\hline\hline
Location of observation & Spacecraft mission/Instrument & Source\\ \hline
 Flyby  & Pioneer 10/CPI & \cite{Teegarden1974}\\ 
 & Ulyssses/KET &\cite{Heber2005}\\
 & Voyager 1/TET &\cite{Nndanganeni2018}
 \\ \hline
 Earth orbit & SOHO/EPHIN & \cite{kuehl2013} \\ 
 (well connected) & Ulyssses/KET & \cite{Heber2005}\\ 
 & ISEE 3/ICE& \cite{Moses1987}\\ 
 & Voyager 1/TET &\cite{Nndanganeni2018}
 \\ \hline
 Earth orbit& ISEE 3/ICE & \cite{Moses1987}\\ 
 (badly connected) && \\\hline
\end{tabular} 
\\[10pt]
\label{tab:spectral_data}
\end{table*}

\section{Calculating Differential Intensities}
\label{sec:diff_intensities}

\begin{figure}
    \centering
    \includegraphics[width=0.9\columnwidth]{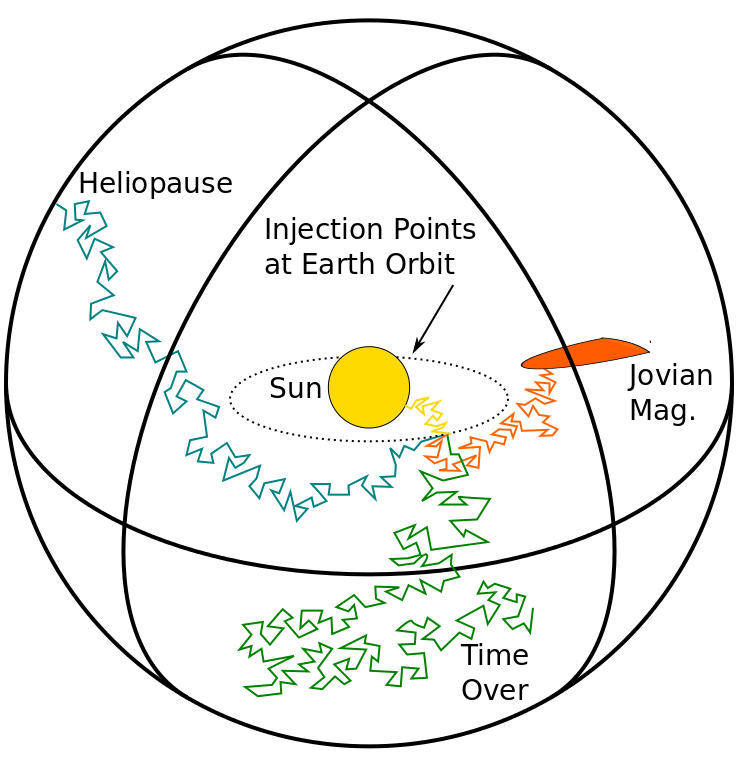}
    \caption{Sketch of the simulation setup as detailed in \cite{Dunzlaff2015}. Color coded the four different exit possibilities are shown: The Sun (yellow), the Jovian magnetosphere (orange), the Heliopause (blue) as well as the option the phase space trajectory is 
    terminated
    by reaching a pre defined end time (green). Note that the figure is not to scale.}
    \label{fig:setup}
\end{figure}

For the purpose of this study, the \ac{SDE} code as discussed by \cite{Dunzlaff2009} was utilized, which is based on previous codes by e.g., \citet{strauss2011,strauss2011b}. This SDE solver is written in CUDA to optimise on performance time, and therefore incorporates a simplified analytical approach for both the solar wind velocity as well as for the \ac{HMF}. The solar wind velocity is chosen to be $u_{SW}=400~$km/s and directed radially outwards. The \ac{HMF} is assumed to be Parker-like and included geometrically within the diffusion tensor according to \citet{Burger2000}. Therefore the results of this study are applicable to solar minimum conditions. 

It has been established, from a modelling  \citep[e.g.][]{Potgieter1996,Burger2000,Ferreira2002} as well as from a theoretical perspective \cite[e.g.][]{Bieber1994,EB2010}, that drift effects can be neglected when studying the transport of electrons with energies of a few MeV.
In order to optimize the performance time of the code, the 
energy-independent
approach of \cite{Dunzlaff2015} and \citet{strauss2011} is used to include the parallel and perpendicular mean free paths: $\lambda_{\parallel}$ is normalized to a value of the parallel mean free path $\lambda_0$ at $1$~AU, such that
\begin{equation}
\label{eqn:lambda_par}
\lambda_{\parallel}(r)=\frac{\lambda_0}{2}\left(1+\frac{r}{r_0}\right)
.\end{equation}

This value of the parallel mean free path, together with the particle speed $\nu$, scales the parallel diffusion coefficient as $\kappa_{\parallel}(r)=\nu\lambda_{\parallel}(r)/3$, and the perpendicular diffusion coefficient via the proportional factor $\chi$ such that
\begin{equation}
\label{eqn:chi}
\kappa_{\perp}(r)=\chi\kappa_{\parallel}(r)\mathrm{.}
\end{equation}

Although the above expressions represent an essentially \textit{ad hoc} approach to modelling diffusion parameters, the 
radial dependency
of the parallel diffusion coefficient does reproduce the corresponding behaviour of the quasi-linear theory electron parallel diffusion coefficient employed by \citet{EB2013}, between $1$~and $5$~AU. Perpendicular mean free path expressions from theory, however, can behave in a manner quite different from what is assumed in this study \cite[see, e.g.,][]{shalchibook,EB2015,gammon}. For a preliminary study, however, the above expression for $\kappa_{\perp}$ should provide a reasonable approximation that meets the requirements for the scientific tasks investigated herein. 
This assumption is confirmed later within this study as the approach leads to reasonable results for the limited energy range considered.

As described in great detail by e. g. \cite{Kloeden2011}, the stochastic nature of diffusion is treated within the \ac{SDE} method as a Wiener process $dW_t=\zeta\sqrt{dt}$ with $\zeta$ being a vector of Gaussian distributed random numbers. The resulting set of four integral equations, as they are derived by \citet{strauss2011}, is solved iteratively by applying the Euler-Maruyama scheme. This leads to a random walk type solution which is terminated if a spatial or temporal boundary is reached, as shown in Fig.~\ref{fig:setup} for various possible exit positions. A time-backward phase-space trajectory can either terminate at the assumed position of the heliopause (blue line), the Jovian magnetosphere (orange line), or after a pre-specified number of steps have been made without an encounter with either of the formerly mentioned structures (green line). As discussed in Sec.~\ref{ssec:jovian:source}, in the low-MeV energy range Jovian electrons dominate the spectrum. The solar population therefore can be assumed to be negligible during quiet times, and thus trajectories exiting at the Sun are discarded. Galactic electrons on the other hand are considered via the local interstellar spectrum according to \cite{Potgieter2017}, although the transport of the Galactic population into the inner heliosphere is much more sensitive to drift effect, which are not considered here. 

Instead of solving Parker's \ac{TPE} directly to 
obtain
a time dependent distribution function $f(\vec{r},t)$ covering the whole phase space, the \ac{SDE} method provides a chain of point-like solutions both in space and time. The solutions must be sampled at a large number of phase-space points to approximate a spatial solution covering the phase space of interest. In order for the computational results to be comparable with spacecraft data, the time-backward setup is solved as derived and discussed by e.g. \cite{Kopp2012}. A comparison between the the time-forward and time-backward setup, made using a simpler 1D model, can be found in the review by \citet{Strauss2017}.



The left panel of Fig.~\ref{fig:influence_spectrum_flux} shows 
histograms 
of the so called exit energy \footnote{Note that the term exit energy refers to a time-backward simulation setup. Therefor the exit energy describes the energy the particle would have been emitted with at its source - which is in the case of this study the Jovian magnetosphere.} (i.e. the energy of which pseudo-particles leave the computational domain)
at times of good magnetic connection between the Jovian magnetosphere and the observational point, in contrast to Fig.~\ref{fig:inf_spectrum_flux_6MeV_worst}, illustrating the energy distribution for an observational point in opposition to the Jovian source. The distributions in Fig.~\ref{fig:inf_spectrum_flux_6MeV_worst} are much broader. This of course is in agreement with Fig.~\ref{fig:inf_spectrum_flux_6MeV_best} being dominated by the much more efficient parallel diffusion along the nominal Parker spirals, while pseudo-particles reaching the Jovian magnetosphere in Fig.~\ref{fig:inf_spectrum_flux_6MeV_worst} must do so via the more inefficient perpendicular diffusion process. These particles suffer much more scattering, leading to a more diffuse distribution.


Taking the integral distributions (dashed lines) into account it shows that these extreme particle trajectories 
corresponding to high exit energies
contribute very little to the total differential intensity. In case of of good connection Fig.~\ref{fig:inf_spectrum_flux_6MeV_best} suggest that the pseudo-particles with exit energies below 10 MeV contribute about 90 \% to the total differential intensity whereas Fig.~\ref{fig:inf_spectrum_flux_6MeV_worst} shows that in case of bad connection they still make up about 80 \%. Therefore, we conclude that although the trajectory of each pseudo-particle is (mathematically) equally likely, they do not contribute equally towards the differential intensity. This is equivalent to stating that different pseudo-particles represent different amounts of physical particles corresponding to their physical significance.

The question now becomes: If each pseudo-particle represents a different number of real particles, why should we weigh the pseudo-particles equally when calculating physical quantities 
, such as the residence time,
from their distribution?

\subsection{Model Parameter Dependencies}
\label{ssec:corotation}

\begin{figure}
            \centering 
            \includegraphics[width=\columnwidth]{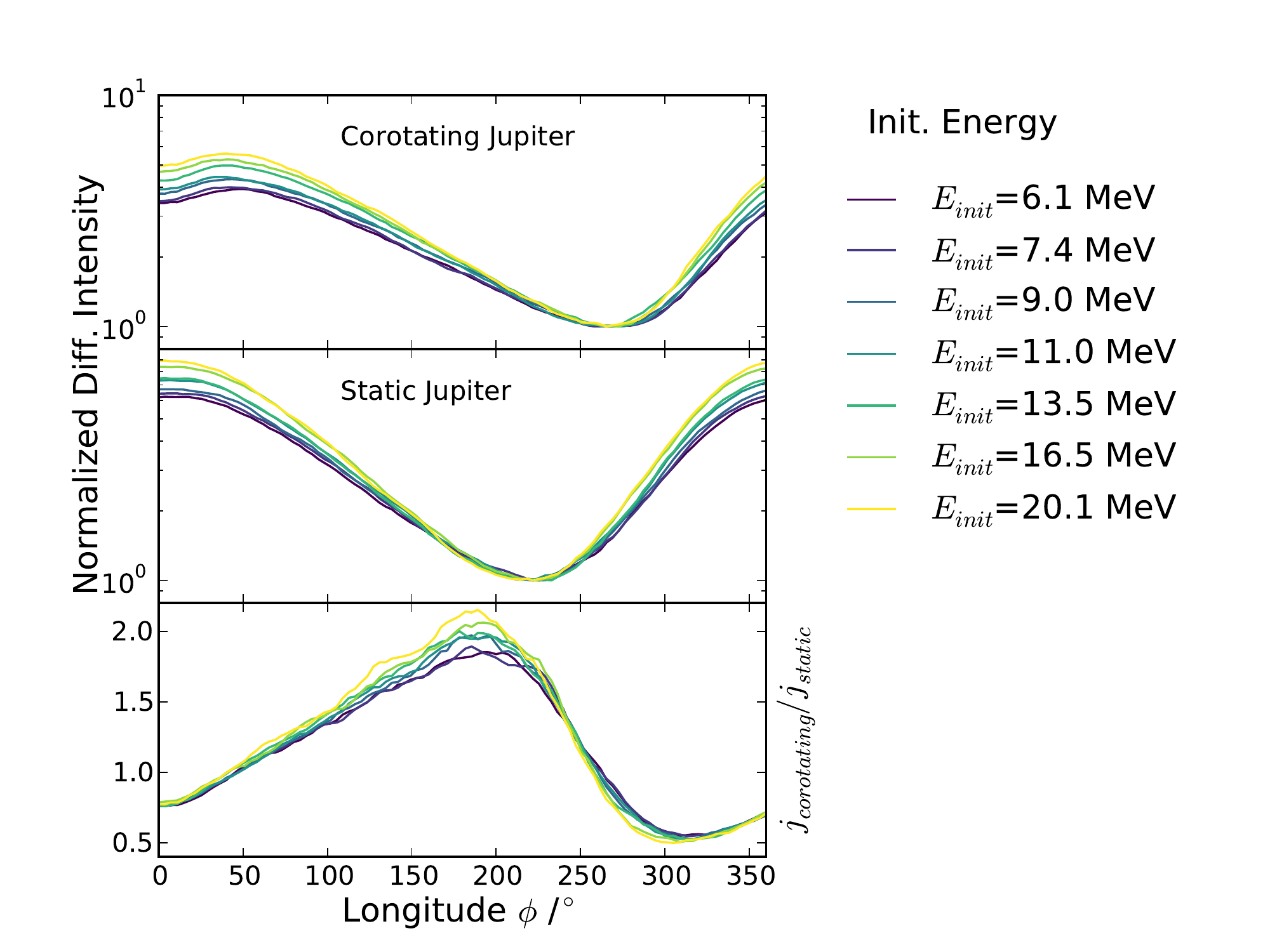}
   \caption{Differential intensities of Jovian electrons for different initial energies $E^{init}$. The top panel shows the longitudinal variation due to varying connection for a co-rotating Jovian source, while the middle panel shows the case of a static Jupiter. The bottom panel shows the ratio of the results of the two approaches.}
    \label{fig:influence_corotation_flux}
\end{figure}


A major aspect of modelling physical processes is to simplify the computational setup in order to save resources and time without affecting the physical validity of the results. Therefore the variability of our results was tested against various model set-ups in order to find the most optimal modelling scenario.

The choice of time increment $\Delta s$ has the most obvious influence on the simulation results due to its association with the Wiener process and hence with the mathematical representation of diffusion. Both the effects of the size 
of the time increments
and the 
have been tested. 
For $\Delta s$ a high sensitivity to the magnetic connection was observed as expected since the time increment (which is related to the adiabatic energy changes) 
influences the duration of the random walk. Thereby it is important to note that the effect is more significant for observation points of good magnetic connections. As the trajectories for bad connections are much more diffusion dominated the effect of the size of the time increment is more likely to be cancelled out by the number of timesteps. For good connections, however, the trajectories are much more dominated by the geometry of the underlying \ac{HMF}. Therefor larger time increments cause larger abbreviations and subsequently longer simulation times. 
It was found that the simulated differential intensity converges for values of $\Delta s \leq 0.001$ in program units (about $0.004~$days). The numerical dependence on the 
maximal duration of simulation
appears to be very weak (above a certain limit), with a required 
duration
of about $700$ to $800$ 
program units
in order to avoid deviations of more than $20~$. This leads to a total time of simulations of $\sum\Delta s\ge 300~$days. As the numerical step sizes of the pseudo-particles are ultimately dependent on the choice of the energy transport parameters (which influences the diffusive step size), there is a slight energy dependence on the optimal size of the timestep.

Due to the fact that Jovian electrons originate 
in
and 
populate mainly
the inner heliosphere, it seems natural to raise the question as to whether a model heliosphere could be restricted in size so that calculation results remain unaffected. It was found that results become insensitive to the size of the model heliosphere if $R_{HP} \geq 80~$AU is used. For smaller radii the resulting differential intensities increase exponentially, regardless of magnetic connection. Therefore we use a value of $R_{HP} = 120~$AU in order to assure convergence, and, as reported by e. g. \citet{Gurnett2013}, 
this is
the radial distance 
at which
\textit{Voyager} 1 
crossed
the heliopause.

A more physical question is whether a co-rotating coordinate system is necessary.
To keep the setup mathematically as simple as possible, 
the code is designed in a way that
both the Sun and the Jovian source are kept fixed. The position of the observer as the starting point of the time-backward phase space trajectories is calculated in a coordinate system relative to the 
slow orbital
motion of Jupiter. These relative position can be kept static 
during the random walk
or include the further co-rotation of the Jovian source 
by means of an additional longitudinal advection 
velocity $V_{\phi}=\Omega_{Sun}\cdot r$.
Fig. \ref{fig:influence_corotation_flux} shows the calculated Jovian electron intensities, at Earth, for different energies as a function of longitude. Simulations were performed for both a co-rotating 
coordinate system
(top panel) and for the case of a static Jupiter (bottom panel). The bottom panel shows the ratio of these two scenarios and proves that the ratio between the differential intensities for a co-rotating and a static approach appears to be significant. We conclude that Jovian co-rotation must be included in the model to produce reasonable results.  We concluded that the co-rotation of the Jovian source is an important effect to incorporate into the numerical model.

\subsection{Deriving an estimation for the mean free paths}
\label{ssec:mean_free_paths}

\begin{figure*}
        \centering
        \begin{subfigure}[b]{\columnwidth}  
    \centering
    \includegraphics[width=0.9\columnwidth]{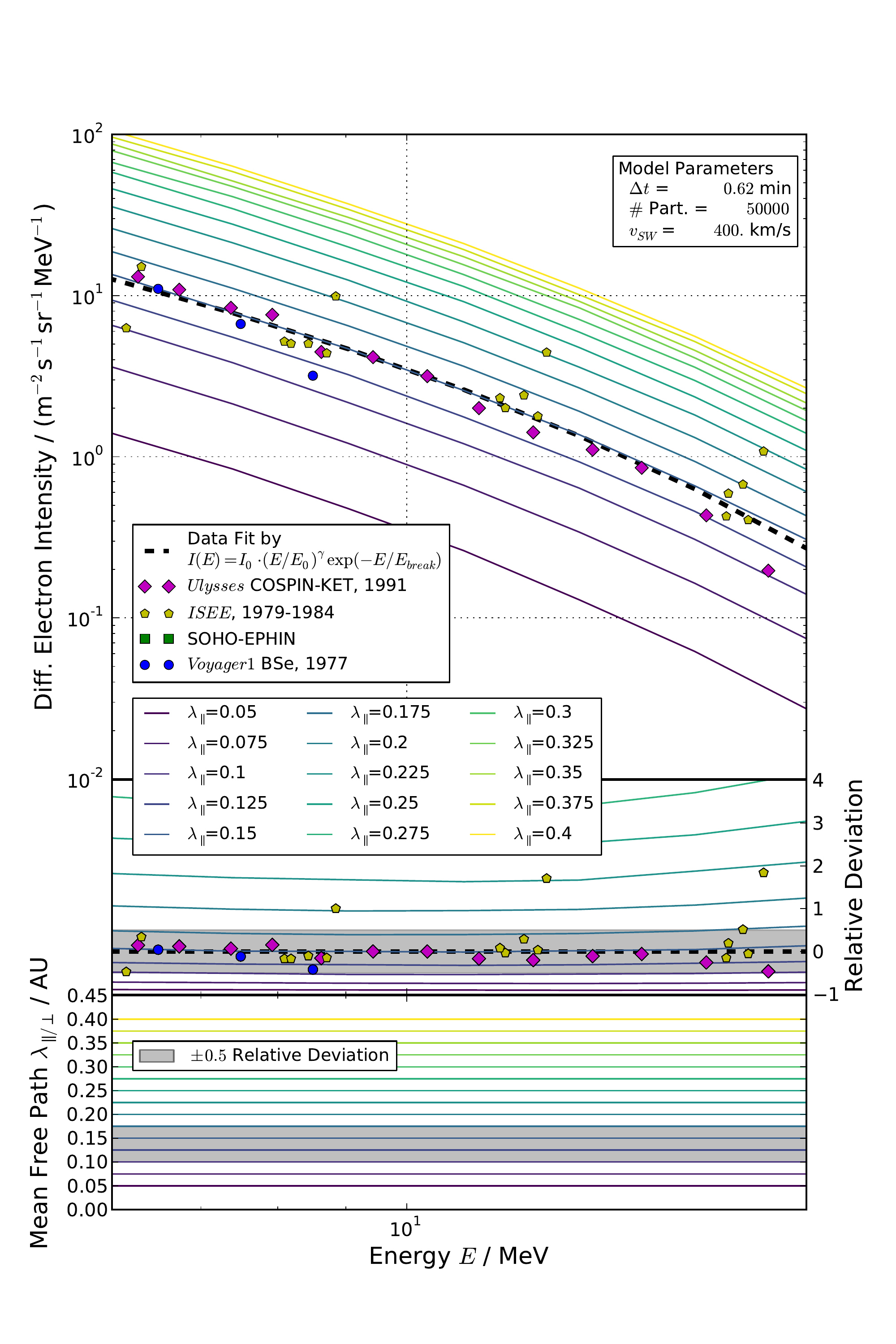}
    \caption{Simulated Jovian electron spectra at the longitudinal point of best effective magnetic connection for different values of $\lambda_{\parallel}=[0.05,0.4]~$AU. 
    The top panel shows a data fit (dashed line) along with the simulations results and spacecraft data obtained during corresponding time periods. The second panel displays the simulations' relative deviation from the fit. In order to estimate the agreement, the $\pm0.5$ deviation area is given in shaded grey together with the spacecraft data. In the lowest panel the shaded area marks the range of values for $\lambda_{\parallel}$ within the margin of less than $\pm 0.5$ relative deviation.}
    \label{fig:spectrum_mfp_par}
        \end{subfigure} 
        \hfill        
        \begin{subfigure}[b]{\columnwidth}
    \centering
    \includegraphics[width=0.9\columnwidth]{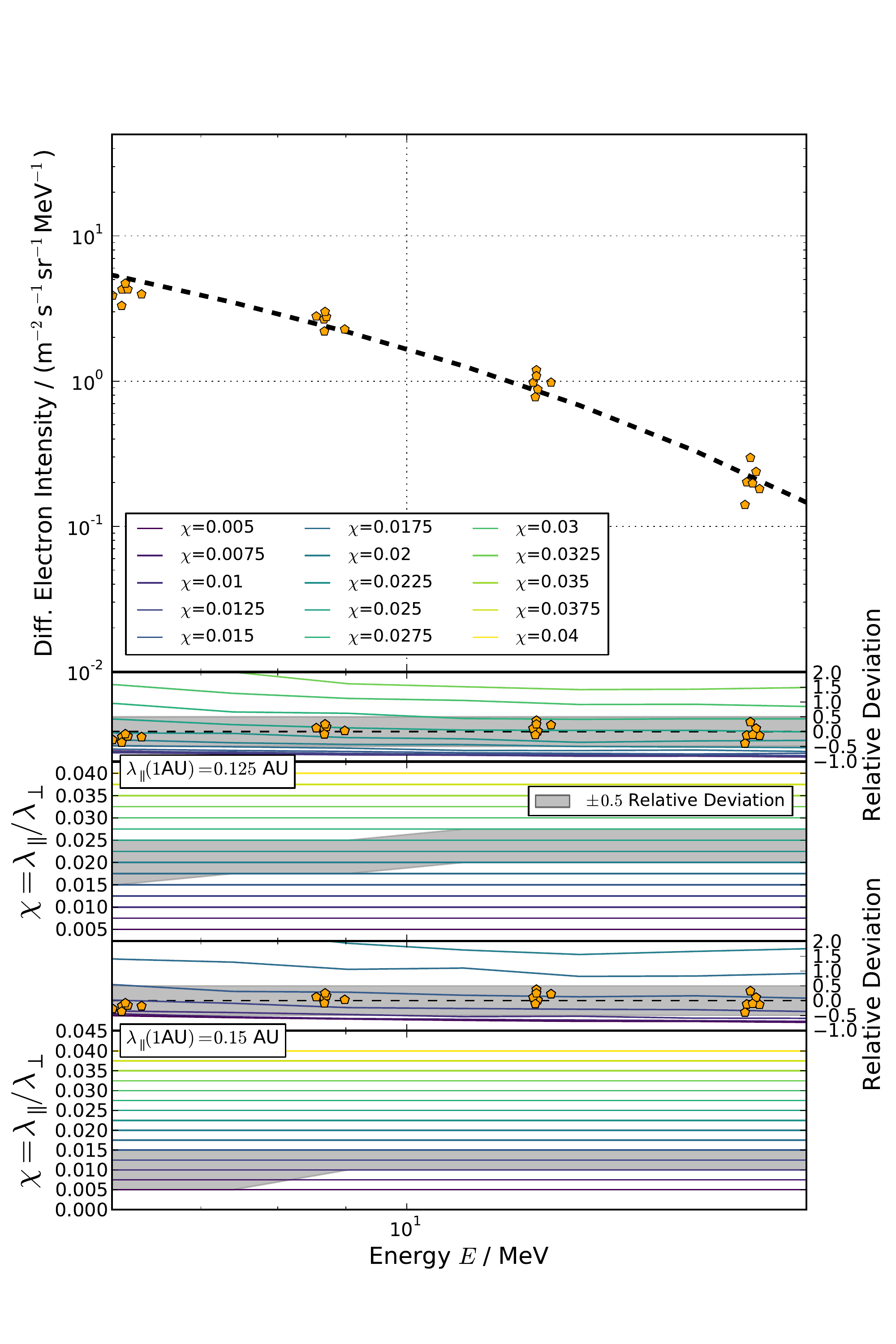}
    \caption{Simulated Jovian electron spectra at the longitudinal point of least effective magnetic connection for different values of $\chi=[0.005,0.045]$. 
    In contrast to Fig.~\ref{fig:spectrum_mfp_par} the simulated spectra are not shown in the upper panel as the simulations are performed utilizing two different values for $\lambda_{\parallel}$. The fit (dashed line) was performed utilizing spacecraft data by \textit{ISEE 3}, the only spectral data for magnetically poorly connected observation times as given by Tab.~\ref{tab:spectral_data}. The second and fourth panels are equivalent to the second of Fig.~\ref{fig:spectrum_mfp_par}, whereas the third and fifth panel correspond to the third of Fig.~\ref{fig:spectrum_mfp_par} showing the possible ranges for $\chi$.}
    \label{fig:spectrum_mfp_perp}
        \end{subfigure}
        \caption{Simulated Jovian electron spectra used to optimize the values of $\lambda_{\parallel}$ and $\chi$. The lower panels show the corresponding relative deviations 
        from the fit (dashed lines)
        and the resulting best fit behaviour for both quantities, 
        giving an estimation of the energy dependency of $\lambda_{\parallel/\perp}$.}
    \label{fig:mean_free_paths_data}   
\end{figure*}

Fig.~\ref{fig:influence_spectrum_flux} shows that only lower-energy pseudo particles (with exit energies near the initial energy) contribute significantly to particle intensity at Earth. As this energy range only covers an order of magnitude, the energy dependence of the mean free paths both parallel and perpendicular can be neglected for the means of this study. Note, however, that the resulting diffusion coefficients, which are the parameters used in the model, might still have some energy dependence due to their dependence on the particle speed $\nu$.

Thus, to estimate $\lambda_0$, the Jovian spectral data at Earth orbit as listed in Tab.~\ref{tab:spectral_data} were compared to the results of a parameter study covering the parameter space in question, namely $\lambda_{\parallel}$ and $\lambda_{\perp}$. As already discussed in Sec.~\ref{ssec:test_particles}, the decentral position of the Jovian source allows one to estimate the effectiveness of parallel and perpendicular transport dependent on the observational point's magnetic connection with Jupiter. In order to ensure consistency, $\lambda_{\parallel}$ was determined when investigating the case of best magnetic connection between the observational point and the source, 
before determining $\chi$ which scales $\lambda_{\perp}$ with respect to $\lambda_{\parallel}$ according to Eqn.~(\ref{eqn:chi}) and therefore dominates the results in case of magnetic opposition.

The upper panel of Fig.~\ref{fig:spectrum_mfp_par} shows  the available Jovian electron spectral data at Earth orbit during times of good magnetic connection. For a more detailed discussion with regards to the Jovian source spectrum see \cite{vogt2018} or the individual publications listed in Tab.~\ref{tab:spectral_data}. The black dashed line marks the shape of the Jovian source as fitted to the Earth orbit data in order to have a measure to define the deviation. Color coded in all three panels are the results of the simulations with varying parallel mean free paths covering the range of $\lambda_{\parallel}=[0.05~$AU$,0.4~$AU$]$ in steps of $0.05~$AU, and a value of $\chi=0.01$ in agreement with the suggestions by \cite{Palmer1982} as well as \citet{Bieber1994} and succesfully implemented in previous studies by \citep[e.g.][amongst others]{strauss2011b,vogt2018} is used. The results are shown in the two upper panels of Fig.~\ref{fig:spectrum_mfp_par}; the top panel depicting the simulated intensities whereas the middle panel shows their relative deviation from the nominal spectrum provided by the fit. The grey area marks the range of a $\pm 0.5$ relative deviation 
in order to give an impression on the reliability of the results. As the errors related to the data appeared to be too small to be significant and it is way beyond the scope of this work to re-examine them, the choice was made to display the simulation results with respect to their relative deviation from the fit.
In the bottom panel, the area between the lowest and the highest value of $\lambda_{\parallel}$, leading to simulation results within this $\pm 0.5$ relative deviation, is again marked in grey. 
Thereby it provides an estimation of the energy dependency of $\lambda_{\parallel/\perp}$.
The energy range is chosen in order to cover the range of exit energies contributing to the total differential intensity for an inititial energy of 6 MeV according to Fig.~\ref{fig:influence_spectrum_flux}. As it appears a value of $\lambda_{\parallel}=0.15~$AU seems to match the observations best over the whole energy range investigated herein.
This is an additional motivation for our choice of an energy-independent mean free path.

Fig.~\ref{fig:spectrum_mfp_perp} shows the relative deviation of simulations with varying $\chi=[0.00,0.04]$ and for two values for $\lambda_{\parallel}=[0.125,0.15]~$AU. 
The uppermost panel of Fig~\ref{fig:spectrum_mfp_perp} shows the spectral data at Earth orbit according to Tab.~\ref{tab:spectral_data},
specifically obtained during times of bad magnetic connection between the spacecraft and the Jovian magnetosphere. Similar to Fig.~\ref{fig:spectrum_mfp_par}, the dashed line indicates the spectral shape of the source when fitted to the Earth orbit data. In this case the data are presumably dominated by the effects of perpendicular diffusion. As already discussed above, larger values of $\lambda_{\parallel}$ lead to more effective diffusion and therefore to increased differential intensities at the observational point. This relation is also present in the results depicted by Fig.~\ref{fig:spectrum_mfp_perp}, as higher values of $\lambda_{\parallel}$ demand smaller values of $\chi$ in order to fit the data. 
Although the estimation of parallel and perpendicular mean free paths based on turbulence theory is still an ongoing topic of research, previous studies suggest values between $0.09$ and $0.3~$AU for $\lambda_{\parallel}$ \citep{Palmer1982,Bieber1994} which were tested successfully for electrons in the inner heliosphere such by studies such as \citep[e. g.][amongst others]{Ferreira2003, Ferreira2005,strauss2011, Droege2016}.
Although values of
$\lambda_{\parallel}=0.175~$AU 
themselves would fit well within this range, they
would demand 
values of $\chi$
out of a realistic range \citet[see e.g.][amongst others]{Palmer1982,Bieber1994,strauss2011,Droege2016}
and therefor are considered as unrealistic.

\begin{table}
\centering
\caption{The computational and physical parameters used for the  code (when not stated otherwise).} 
\begin{tabular}{l l}
\hline\hline
\multicolumn{2}{l}{Computational Parameters}\\ \hline
\# Trajectories & 50000 \\
$\Delta t$ & 0.0001\\
$T_{End}$ & 800\\\hline
\multicolumn{2}{l}{Physical Parameters}\\ \hline
$R_{HP}$ & $120~$AU \\
$u_{SW}$ & $400~$km/s\\
$\lambda_{\parallel}(1~$AU$)$ & $0.15~$AU \\
$\chi=\lambda_{\perp}/\lambda_{\parallel}$ & $0.0125$\\
$E^{init}$ & $6~$MeV\\
\end{tabular} 
\\[10pt]
\label{tab:simulation_parameters}
\end{table}

Taking the best fitting $\chi$ for  
$\lambda_{\parallel}=0.15~$AU 
(as the best fitting value for $\lambda_{\parallel}$ according to Fig.~\ref{fig:spectrum_mfp_par})
into account, a value of $\chi=0.0125$ seems to be justifiable within the margin of error for the energy range of interest. This value for the parallel mean free path is well within the \citet{Palmer1982} consensus range, and reasonable for low-energy electrons \cite[e.g.][]{Bieber1994}. The value for $\chi$ is somewhat smaller than the range expected from \citet{Palmer1982}, where $0.02 \leq \chi \leq 0.08$, but well within the range of values commonly used in numerical modulation studies \cite[see, e.g.,][]{Ferreira2001,EB2013,Nndanganeni2018}. Therefore, unless otherwise indicated, these parameters were utilized throughout this study as summarized in Tab.~\ref{tab:simulation_parameters}.

\subsection{Interpretation of phase space trajectories}
\label{ssec:phase_space_traj}

\begin{figure*}
        \centering
        \begin{subfigure}[b]{\columnwidth}
        \centering
     \includegraphics[width=\columnwidth]{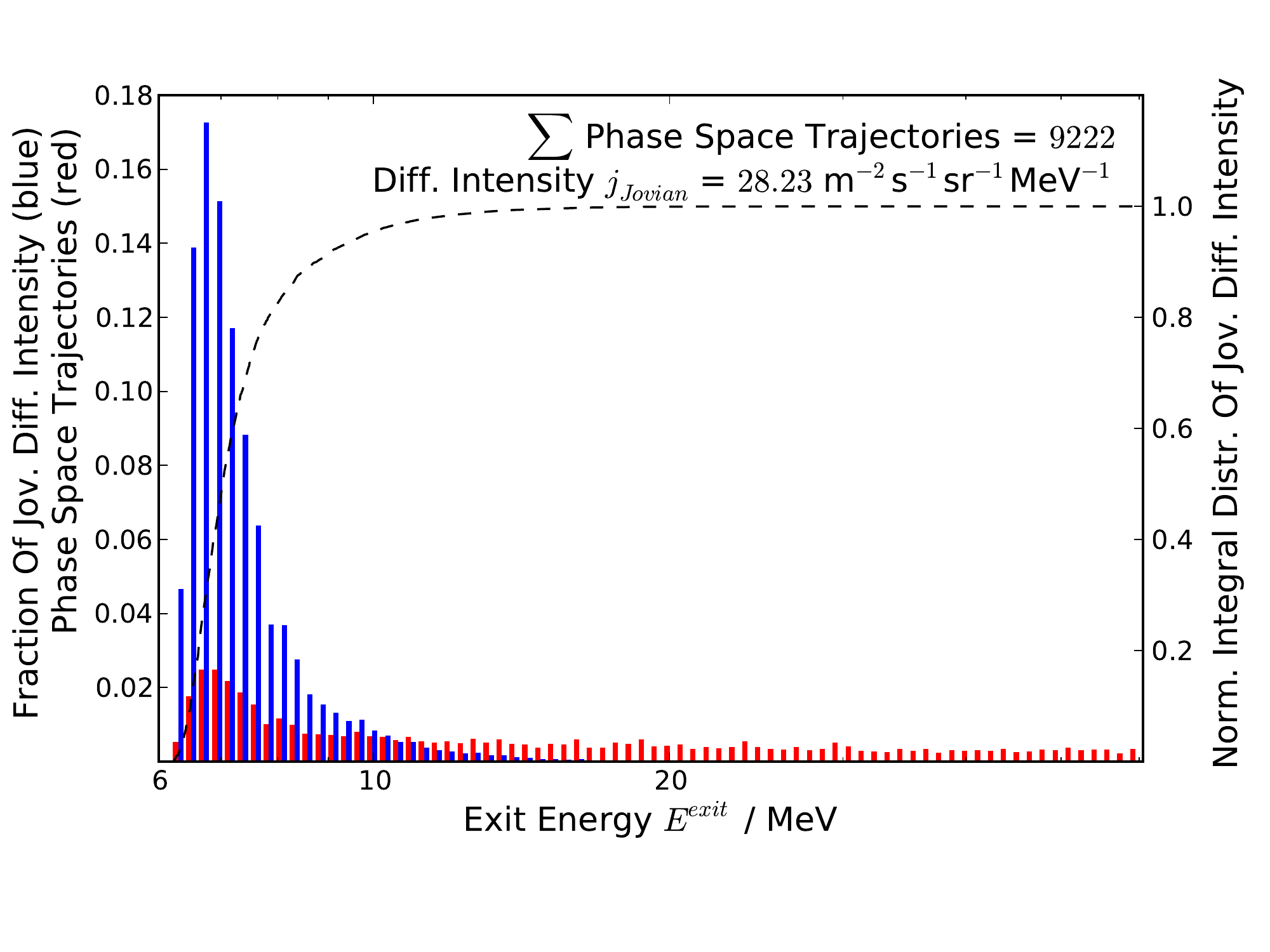}
            \caption
            {\normalsize Distribution of exit energies $E^{exit}_i$, simulated with an observational point at Earth orbit, magnetically well connected to the Jovian source}
 \label{fig:inf_spectrum_flux_6MeV_best}
        \end{subfigure}
        \hfill
        \begin{subfigure}[b]{\columnwidth}  
     \includegraphics[width=\columnwidth]{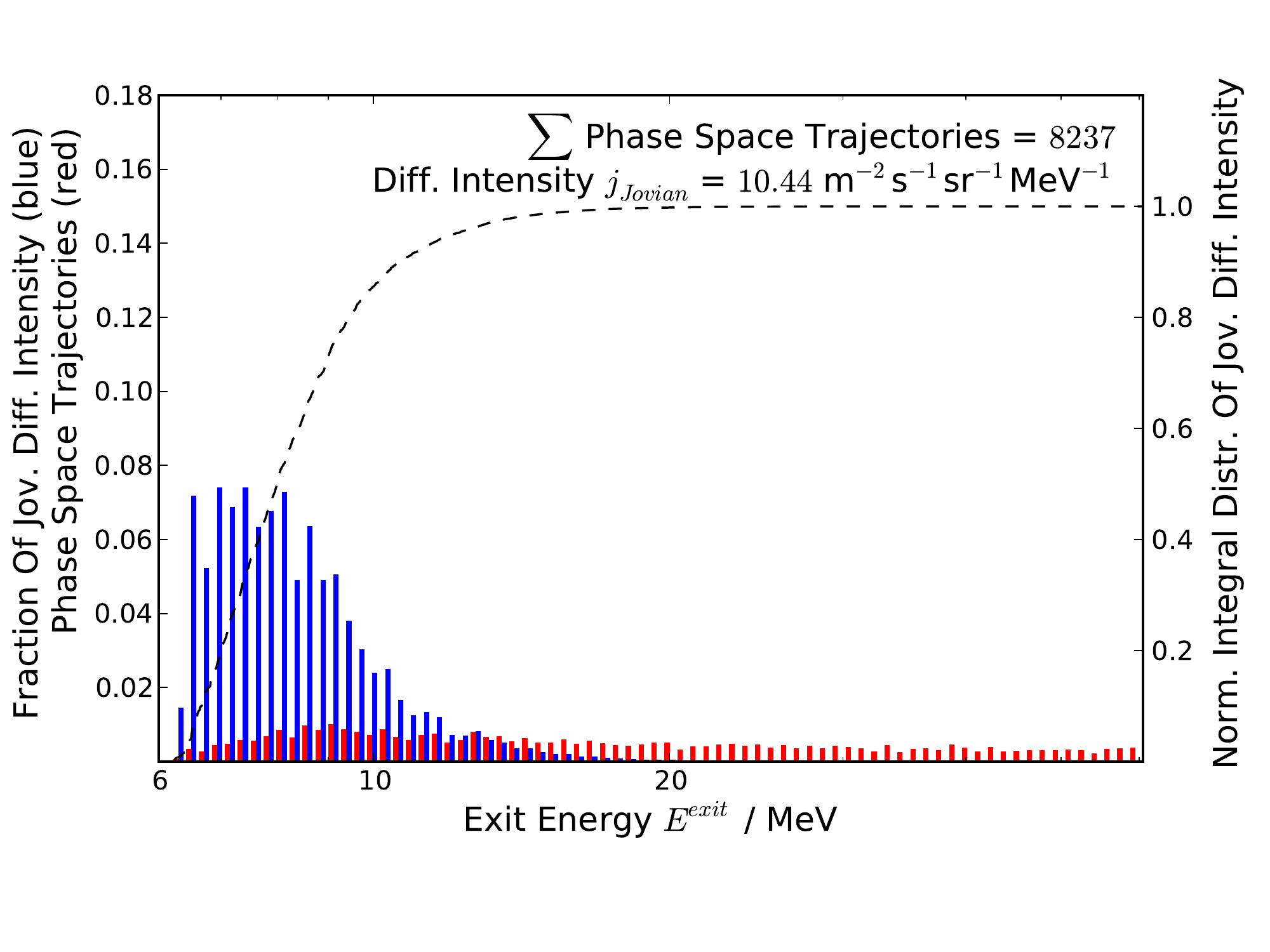}
            \caption
            {\normalsize Distribution of exit energies $E^{exit}_i$ simulated with an observational point at Earth orbit magnetically poorly connected to the Jovian source}
            \label{fig:inf_spectrum_flux_6MeV_worst}
        \end{subfigure}
    \caption{Binned distributions of the exit energies $E^{exit}_i$ (red) and the binned contribution of each energy point to the total differential intensity, $j_{jov}(E^{exit}_i)$ 
    (blue), for an initial energy of $E^{init}_i=6~$MeV.
    . The dashed lines show the integral distribution of the trajectories' contribution to the total differential intensity as shown in blue.} 
    \label{fig:influence_spectrum_flux}
\end{figure*}

As already pointed out above, the phase space trajectories obtained by the \ac{SDE} solver are mathematical solutions and, in contrast to a physical interpretation, each phase space trajectory is equally probable. However, not each phase space trajectory represent the same number of physical particles. Although they are often referred to as pseudo-particles, this term is slightly misleading because the phase space trajectories represent the evolution of the particle density distribution $f$ along a curve through the heliosphere and has no connection with the trajectory of actual charged particles in a turbulent plasma.

The \ac{TPE} is solved by integrating the \acp{SDE} via the time-backward Euler-Maruyama scheme which leads to an increase of the pseudo-particle's energy $E_i$ due to the inverse adiabatic processes (adiabatic energy losses treated in a time-backwards fashion). Applying therefore the source spectrum $j_{jov}(E)$ as the boundary weight \citep[see, e. g.,][]{strauss2011,Kopp2012} leads to the expression 

\begin{equation}
\label{eqn:diff_intensities}
j(r^0,E^0)=\frac{\sum^N_{i=1}j_{jov}(E^{exit}_i)}{N}
\end{equation}

with the distribution function $f$ as the solution of the TPE related to the differential intensity $j=P^{2}f$, with $P=pc/q$ being the particle rigidity, depending on the momentum $p$ and the charge $q$. 

Eqn.~(\ref{eqn:diff_intensities}) can also be derived by calculating the Green's function $G(x_i,s=T)$ for any given boundary or initial condition in order to solve the convolution \cite[see, e.g.,][]{Pei2010}
\begin{equation}
\label{eqn:conv_sde}
    f(x,T)=\int_0^T\int_x G(x',t)f_b(x',t)dx'dt
\end{equation}
with $f_b(x',t)$ denoting the boundary value. As discussed by \cite{Strauss2017}, neglecting the initial condition simplifies Eqn.~(\ref{eqn:conv_sde}) to
\begin{equation}
\label{eqn:conv_sde_tpe}
    f(\vec{x}_0,t_0)=\int_0^{t_0}\int_{\vec{x}\in\Omega_b}f_b(\vec{x},t)\rho(\vec{x},t)d\Omega dt
\end{equation}

with $\vec{x}=(r,E)$, $f_b(\vec{x},t)$ representing the boundary condition (i.e. the particle population's source distribution), $\rho(\vec{x},t)$ being the conditional probability density and $f(\vec{x}_0,t_0)$ the distribution at the observational point. Focusing on steady state solutions where $t\rightarrow\infty\Rightarrow\rho(\vec{x}_i,t):\rightarrow\rho(\vec{x}_i)$, as well as calculating $f_i(\vec{x}_0,t_0)$ for each phase space trajectory $\rho_i$ individually reducing the spatial boundary of the integration domain $\Omega_b$ to the exit position $r^{exit}$ as well as the range of the energy coordinate to $E^{exit}$, then leads to the expression as given by Eqn.~(\ref{eqn:diff_intensities}).

 Fig.~\ref{fig:influence_spectrum_flux} shows the probability density (binned distribution of pseudo-particles, $N_i/N$; red) and the elements of Eqn.~(\ref{eqn:diff_intensities}) ($j_{jov}(E^{exit}_i)$; blue) for an initial energy of $E_0=6~$Mev at an observational point at Earth. Whereas each pseudo-particle is equally weighted in the calculation, they may contribute very differently to the total differential intensity; from the figure it is evident that pseudo-particles may reach the Jovian magnetosphere with several tens of MeV (which implies significant energy loss in the normal time-forward scenario). However, the blue histogram hints that only pseudo-particles with up to 10 MeV contribute 
 significantly
 to the total differential intensity. The integral distributions 
 (dashed line)
 for both cases further support this assumption 
 as they converge roughly at the exit energies for which the fractional contribution of the differential intensity gets lower than the fractional contribution of pseudo-particles.
 Looking at just the distribution of the pseudo-particles may therefore be misleading if we are interested in deriving physical quantities from the SDE method. It is important to note that, even if some pseudo-particles contribute very little to the differential intensity, they cannot be 
 disregarded
 in the calculations as they still contribute to the denominator of Eqn.(~\ref{eqn:diff_intensities}).

\section{Residence Times}
\label{sec:mean_prop_time}

\begin{figure*}
        \centering
        \begin{subfigure}[b]{\columnwidth}
            \centering
            \includegraphics[width=\columnwidth]{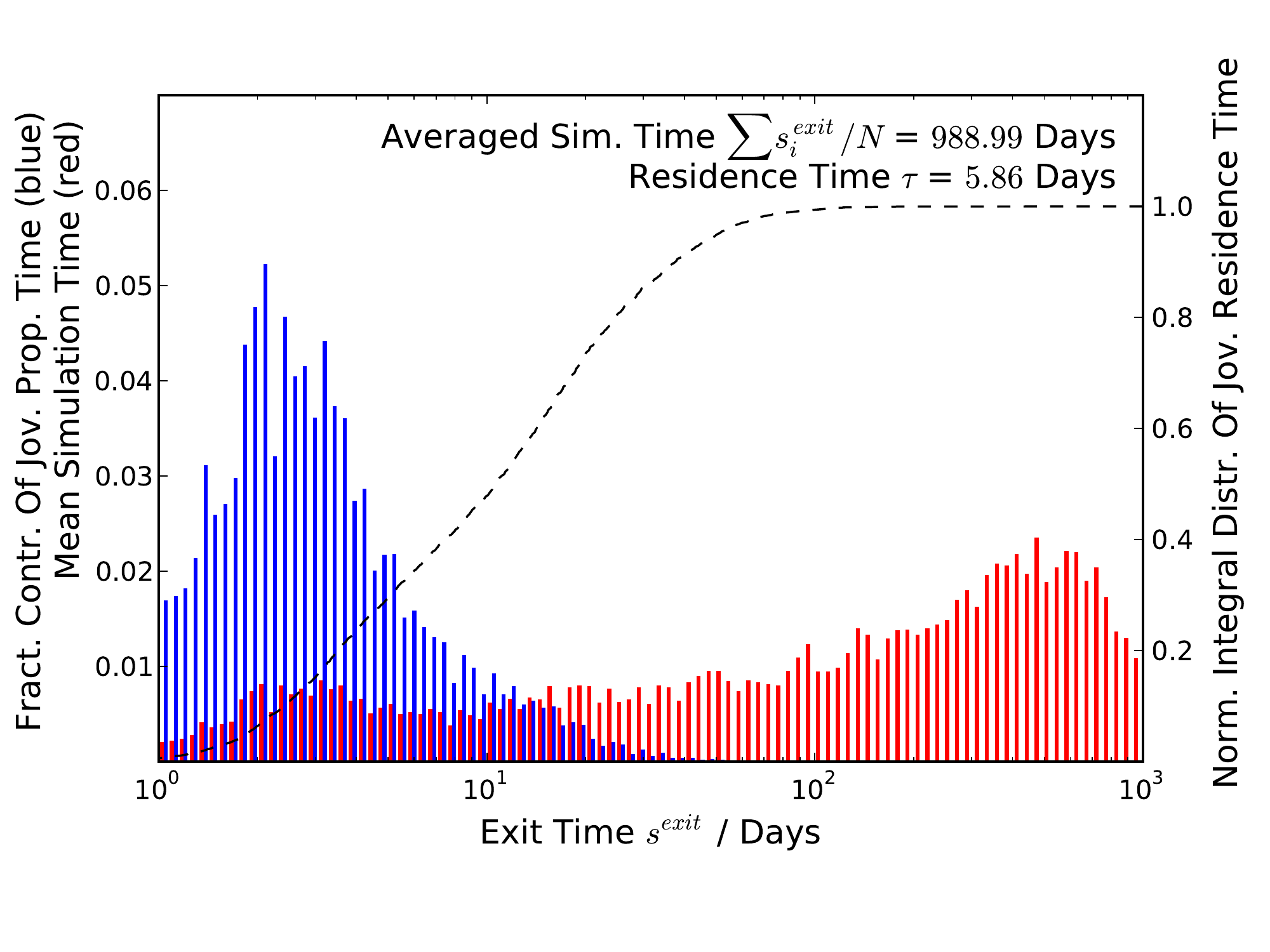}
            \caption
            {\normalsize Distribution of exit times of pseudo-particles for a good magnetic connection between the observational point and the Jovian source.}    
            \label{fig:inf_spectrum_mpt_6MeV_best}
        \end{subfigure}
        \hfill
        \begin{subfigure}[b]{\columnwidth}  
            \centering 
            \includegraphics[width=\columnwidth]{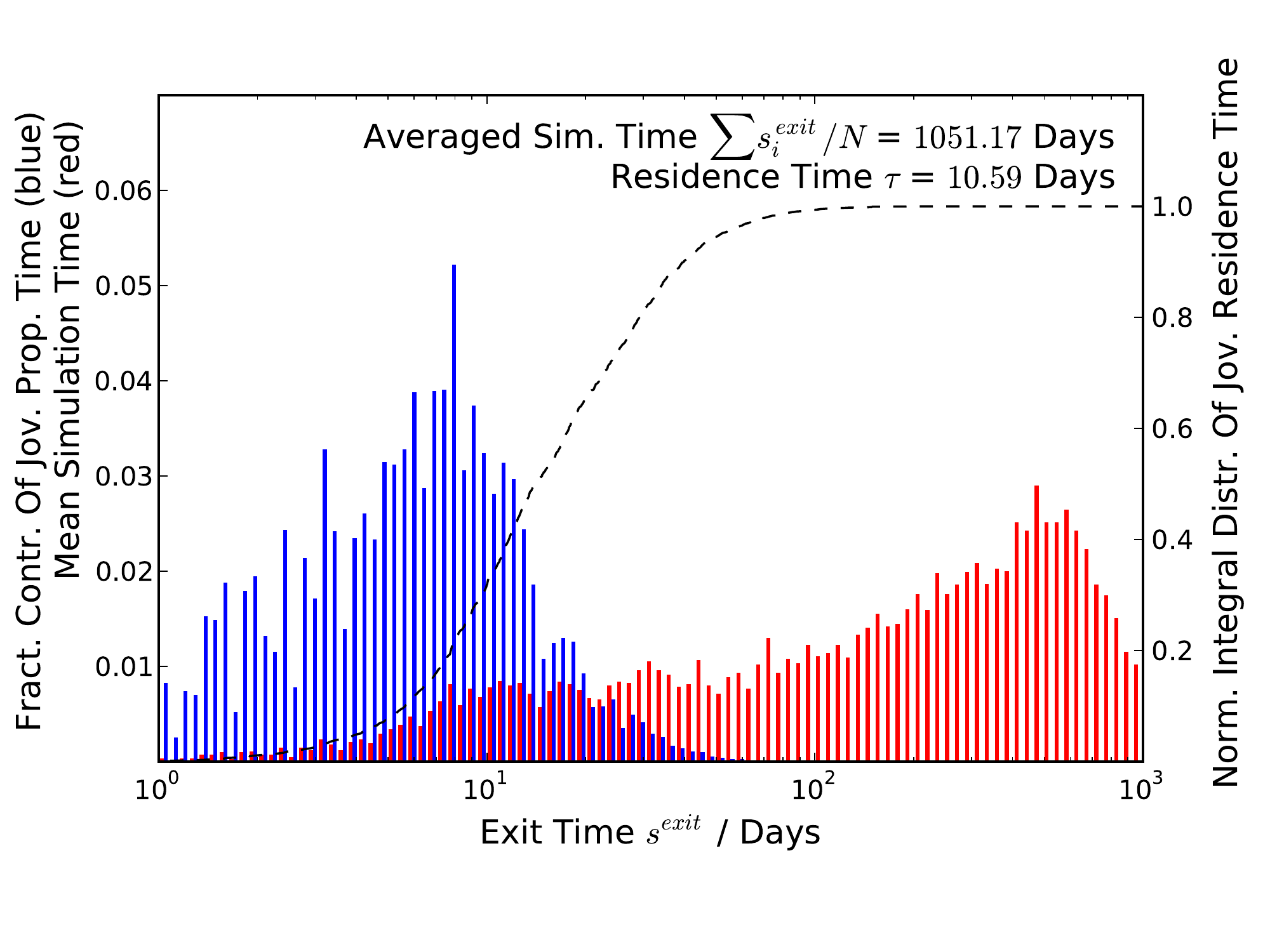}
            \caption[]%
            {\normalsize Similar to the left panel, but now for the case of poor magnetic connection between the observational point and the Jovian source.}    
            \label{fig:inf_spectrum_mpt_6MeV_worst}
        \end{subfigure}
    \caption{Binned distributions of the Jovian electron propagation time using the normal (red) and new (blue) way of defining the probability distribution. The blue distribution thereby shows the weighted exit times, which we refer to as the propagation (residence) time of Jovian electrons. Again the integral distribution of the pseudo-trajectories contribution to $\tau$ is shown by the dashed line. The residence times are obtained via Eqn.~(\ref{eqn:mean_prop_time}). }
    \label{fig:influence_spectrum_mpt}
\end{figure*}  

Because the Euler-Maruyama-Scheme (as well as alternative methods) utilizes a well-defined time increment $ds$ in order to solve the system of integral equations, it is possible to derive a measure $\tau$ as the time that the phase space trajectory takes to connect the source to the observational point. Thereby $\tau$ is calculated as the number of time steps needed for the phase space trajectory $n_i$ multiplied by the time increment $ds$, if the latter time step is assumed to be constant. Generally $\tau$ can be given as $\tau_i=\rvert s_i^{exit}-s_i^0\rvert$, with $s_i^{exit} = n_i\cdot ds$ and $s_i^0$ often being simply the start time of the calculation. Although $\tau_i$ is often referred to as the propagation time or as the residence time too, the more accurate description would be to call $\tau_i$ the trajectory duration or simulation time.

This estimation raises the question of how to interpret the individual phase space trajectories. As discussed in Sec.~\ref{sec:diff_intensities}, phase space trajectories are equally mathematically possible but not equally physically significant. If we now want to derive physical quantities from these trajectories, how should we weigh the resulting distributions? We aim to address this question in the next paragraphs.

\subsection{Numerical Estimation}
\label{ssec:numerical_method_mpt}

Traditionally, the residence or propagation time is assumed to be equivalent to the expectation value of the time, as weighed by a probability density $\rho$,

\begin{equation}
\label{eqn:mean_prop_time_theory}
    \tau = \frac{\int \rho (\vec{x}, t) t dt}{\int \rho (\vec{x},t) dt}.
\end{equation}

Usually, in previous work, $\rho$ is constructed from the SDE solutions as the (normalized) distribution of the pseudo-particles' exit time \citep[see e. g.][]{Florinski2009,Strauss2013}. For 6 MeV Jovian electrons, these distributions are shown in Fig.~\ref{fig:influence_spectrum_mpt}, as the red histogram, for the case of good (left) and bad (right) magnetic connection to the source. From the red distribution, a propagation time of $\sim 550 - 600$ days is calculated, and from the figure itself, we note that most pseudo-particles only reach the observer within $\sim 100 - 1000$ days. This seems very long for relativistic electrons diffusing only a radial distance of $\sim 4$ AU towards Earth. 

An observational estimate to compare with is discussed by \citet{Strauss2013}. Investigating \acp{QTI} \citep[see e. g.][]{McDonald72, Chenette1980} which 
relate
to a diffusion barrier between Earth and Jupiter these authors find that it takes $\approx5$~days after the diffusion barrier has passed Jupiter for the Jovian electrons to be detected again. 

The
numerical
definition of $\tau$ 
according to Eqn.~(\ref{eqn:mean_prop_time_theory})
essentially weighs each pseudo-particle with the same probability. However, we have already seen earlier in this paper that each pseudo-particle does not represent the same number of physical particles. We therefore propose to rather use the distribution of particle density to calculate the propagation time by specifying

\begin{equation}
    \rho (\vec{x},t) = \frac{f(\vec{x},t)}{f_0(\vec{x})}
\end{equation}

which is normalized by the total phase-space density

\begin{equation}
    f_0 (\vec{x}) = \int f(\vec{x},t) dt.
\end{equation}

Thereby $f(\vec{x},t)$ represents the solution of the \ac{TPE} and is calculated
at the exit position $\vec{x}^{exit}$ of the random walk.
Note that $f_0(\vec{x}^{exit})$ cancels in the calculation of $\tau$, which in discrete form reads as

\begin{equation}
\label{eqn:mean_prop_time}
\tau(r^0,E^0) =\frac{\sum^N_{i=1} s(E^{exit}_i)\cdot f(E^{exit}_i)}{\sum^N_{i=1}f(E^{exit}_i)},
\end{equation}

and $s$ is the exit (integration) time of the pseudo-particles. Using this new definition, their weighted contribution to the propagation times are shown in Fig.~\ref{fig:influence_spectrum_mpt} as the blue distribution. 
The parameters used for this simulation are listed in Tab.~\ref{tab:simulation_parameters} as discussed in Sec.~\ref{ssec:mean_free_paths}.
These weighted propagation times are generally much shorter than those obtained by weighing the pseudo-particles equally. 
Similar values as the ones found by this study were only obtained using larger mean free paths \citep[compare e. g.][]{Strauss2013} which would turn out to be unrealistic as discussed in Sec.~\ref{ssec:mean_free_paths} as  our findings are in agreement with prior studies on electron mean free paths such as \citep[][amonst others]{Palmer1982,Bieber1994,Tautz2013} and modelling approaches by \citep[e g.][amongst others]{Potgieter2002,Droege2005,strauss2011}.
By integrating the blue histograms, we find propagation times of $\sim 5 - 11$ days, almost two orders of magnitude shorter than the traditional calculation. Again, the comparison with the red histogram showing the pseudo-particle trajectories' exit times makes this difference comprehensible. The integral distribution of the blue histograms 
(dashed lines)
illustrate that the maximum of the exit times distribution (red) is almost a magnitude higher than the range of convergence. 
Normalized to the value of the residence time, the integral distributions (
\textbf{like} for the differential intensities as shown in Figs.~\ref{sec:diff_intensities}) illustrate how little long exit times corresponding to large exit energies influence the residence times if calculated via Eqn.~(\ref{eqn:mean_prop_time}).
Thus leading to the disparity between the averaged exit or simulation times and our estimations of the residence time equivalent to the differences between the two distributions.

As we have shown and discussed in Sec.~\ref{ssec:phase_space_traj}, 
this is caused by
the diverging physical significance of the phase space trajectories.
As Eqn.~(\ref{eqn:mean_prop_time}) addresses and solves this problem,
 we consider 
 our
 new calculation as being more representative of the propagation time of a physical particle. This is supported by the fact that Eqn.~(\ref{eqn:mean_prop_time}) considers the exit or simulation times according to their representation of physical particles and therefore provides a measure consistent with the total differential intensity. 

\section{Discussion and conclusions}
\label{sec:discussion}

\begin{figure}
            \centering
            \includegraphics[width=\columnwidth]{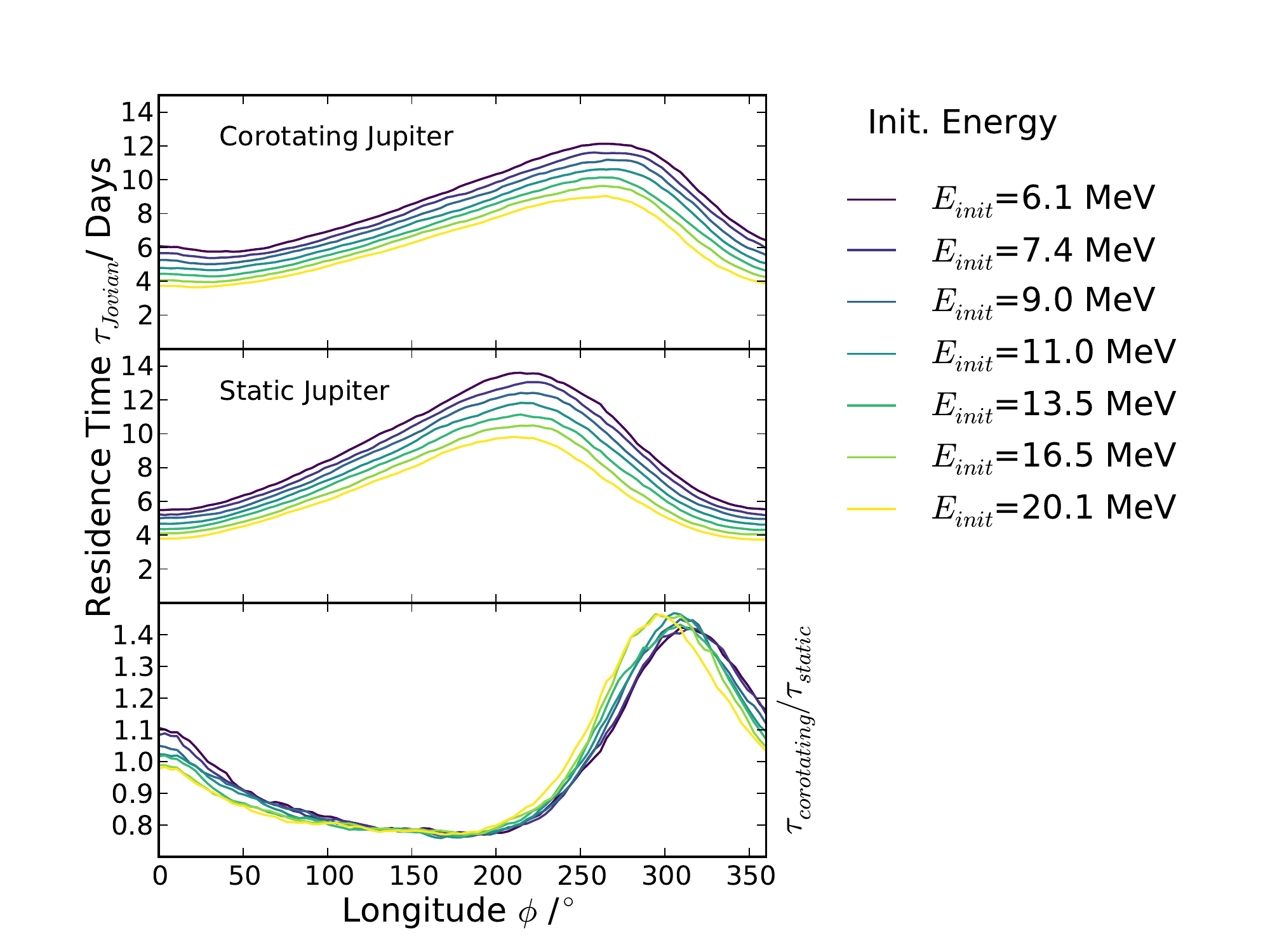}
   \caption{Similar to Fig.~\ref{fig:influence_corotation_flux}, the residence times $(\tau)$ of Jovian electrons are now shown in order to illustrate the influence of co-rotation of the Jovian source as compared to a static source. The bottom panel shows the ratio of the solutions. Different colours indicate different energies.}
            \label{fig:influence_corotation_tau}
\end{figure}

In this paper we have discussed a Jovian electron transport model that was used to study the propagation times (residence times) of 6 MeV electrons.

We have discussed the optimal numerical set-up, including the choice of time step, integration times, and the required size of the numerical domain. We have also shown that taking into account the co-rotation of the Jovian source leads to non-negligible effects. This is an important effect due to its influence on the finite propagation time of Jovian particles from the observer (assumed at Earth, but at different longitudes) to the source. Fig.~\ref{fig:influence_corotation_tau} shows how the calculation of, for example, the propagation time, changes due to the assumption of either a static or a co-rotating source.

We have also used the unique magnetic geometry of the Jovian propagation problem to quantify the appropriate mean free paths. By reproducing the 1 AU intensity spectrum during times of good magnetic connection, we were able to show that $\lambda_{||} \approx 0.1$ AU is appropriate for Jovian electrons in the energy range under consideration. By doing the same during times of bad magnetic connection, we were able to show that a value of $\chi \approx 0.01$ is appropriate. Both these values fall within the range of what is expected from previous studies.

Using the optimal numerical set-up, we calculated the exit times of 6 MeV electrons using the traditional SDE formalism where each pseudo-particle contributes equally to the final results. A value of $\sim 600$ days was found, which, although formally correct, does not seem physically consistent with relativistic particle propagating across a distance of $\sim 4$ AU. Therefore, we propose a new technique to weigh the resulting exit times with the particle intensity, leading to more realistic values of $\sim 6$ days. We propose to term these weighted exit times as the propagation (residence) time of Jovian electrons. The motivation for this weighing is as follows: Although the trajectory of each pseudo-particle is equally probable, each pseudo-particle does not represent the same number of actual particles since the phase-space density of each pseudo-particle is different. 

Especially regarding \acp{CIR} a more realistic measure of Jovian residence times could improve our understanding. As Jovian electron simulations successfully have been applied to this topic by \citep{Kissmann2003,Kissmann2004} the numerical set-up provided by this study provides the opportunity to revisit this approach with further insight. Simulations of residence times could also help to determine how much time Jovian electrons (and other particle populations) spend within structures like \acp{CIR} or magnetic flux tubes. More recently \citep{Daibog2013} highlighted the role of Jovian electrons as test particles within this matter again, suggesting that deviations from their quiet time variations could serve as probes for the inner heliosphere's structure. 

%
\begin{acknowledgements}
This work is based on the research supported in part by the National Research Foundation of South Africa (Grant Number 111731) and the Deutsche Forschungsgemeinschaft (Grant Number FI 706/14). Opinions expressed and conclusions arrived at are those of the author and are not necessarily to be attributed to the NRF or DFG, respectively. KH acknowledges the International Space Science Institute and the supported International Team 464 (ETERNAL).

\end{acknowledgements}
\bibliographystyle{aa}
\bibliography{refs,add}
\end{document}